\documentclass[dvips]{article}
\makeatletter
\newcommand{\singlespacing}{\let\CS=\@currsize\renewcommand{\baselinestretch}
{1.0}\tiny\CS}
\newcommand{\doublespacing}{\let\CS=\@currsize\renewcommand{\baselinestretch}
{1.5}\tiny\CS}
\begin{document}
\textwidth 16cm
\newcommand{\bd}{\begin{document}}
\newcommand{\ed}{\end{document}}
\newcommand{\bc}{\begin{center}}
\newcommand{\ec}{\end{center}}
\newcommand{\bfr}{\begin{flushright}}
\newcommand{\efr}{\end{flushright}}
\newcommand{\lt}{\left}
\newcommand{\rt}{\right}
\newcommand{\vs}{\vspace}
\newcommand{\hs}{\hspace}
\newcommand{\beq}{\begin{equation}}
\newcommand{\eeq}{\end{equation}}
\newcommand{\lb}{\linebreak}
\newcommand{\pb}{\pagebreak}
\newcommand{\mb}{\makebox}
\newcommand{\fb}{\framebox}
\newcommand{\mc}{\multicolumn}
\newcommand{\ben}{\begin{enumerate}}
\newcommand{\een}{\end{enumerate}}
\newcommand{\bit}{\begin{itemize}}
\newcommand{\eit}{\end{itemize}}
\newcommand{\ol}{\overline}
\newcommand{\un}{\underline}
\newcommand{\lefq}{\lefteqn}
\newcommand{\ba}{\begin{array}}
\newcommand{\ea}{\end{array}}
\newcommand{\beqa}{\begin{eqnarray}}
\newcommand{\eeqa}{\end{eqnarray}}
\newcommand{\beqas}{\begin{eqnarray*}}
\newcommand{\eeqas}{\end{eqnarray*}}
\newcommand{\bfg}{\begin{figure}}
\newcommand{\efg}{\end{figure}}
\newcommand{\bds}{\begin{displaymath}}
\newcommand{\eds}{\end{displaymath}}
\newcommand{\btb}{\begin{tabbing}}
\newcommand{\etb}{\end{tabbing}}
\newcommand{\para}{\parallel}
\newcommand{\pad}{\partial}
\newcommand{\nn}{\nonumber}
\newcommand{\la}{\leftarrow}
\newcommand{\ra}{\rightarrow}
\newcommand{\lgla}{\longleftarrow}
\newcommand{\lgra}{\longrightarrow}
\newcommand{\La}{\Leftarrow}\newcommand{\Ra}{\Rightarrow}
\newcommand{\Lra}{\Leftrightarrow}
\newcommand{\Lgla}{\Longleftarrow}
\newcommand{\Lgra}{\Longrightarrow}
\newcommand{\bm}{\boldmath}
\newcommand{\lan}{\langle}
\newcommand{\ran}{\rangle}
\renewcommand{\a}{\alpha}
\renewcommand{\b}{\beta}
\newcommand{\g}{\gamma}
\newcommand{\G}{\Gamma}
\renewcommand{\d}{\delta}
\newcommand{\eps}{\epsilon}
\newcommand{\Th}{\Theta}
\newcommand{\s}{\sigma}
\newcommand{\lam}{\lambda}
\newcommand{\D}{\Delta}
\newcommand{\vare}{\varepsilon}
\newcommand{\pr}{\prime}
\newcommand{\ro}{\rho}
\newcommand{\nab}{\nabla}
\newcommand{\m}{\mu}
\newcommand{\n}{\nu}
\newcommand{\Sg}{\Sigma}
\newcommand{\p}{\pi}
\newcommand{\R}{I\!\!R}
\newcommand{\om}{\omega}
\newcommand{\Om}{\Omega}
\newcommand{\ze}{\zeta}
\newcommand{\vart}{\vartheta}
\newcommand{\tri}{\triangle}
\newcommand{\f}{\frac}
\newcommand{\iny}{\infty}
\newcommand{\pro}{\propto}

\bc
{\Large \bf Generation of exactly solvable non-Hermitian potentials \\
with real energies}
\ec

\vs{1cm}

\bc
{\it Anjana Sinha {\footnote {e-mail : anjana23@rediffmail.com}} \\
Dept. of Applied Mathematics \\
University of Calcutta \\
92, A.P.C. Road \\
Kolkata - 700 009 }
\ec

\vs{.5cm}

\bc
{and}
\ec

\vs{.5cm}

\bc
{\it Pinaki Roy {\footnote {e-mail : pinaki@isical.ac.in}} \\
Physics \& Applied Mathematics Unit \\
Indian Statistical Institute \\
203 B.T. Road \\
Kolkata - 700 108 }
\ec

\vs{2cm}

\bc
{\un{Abstract}}
\ec

\vs{1cm}

A series of exactly solvable non-trivial complex potentials (possessing 
real spectra) are generated by applying the Darboux transformation 
to the excited eigenstates of a non-Hermitian potential $V(x)$.
This method yields an infinite number of
non-trivial partner potentials, defined over the whole real line,  whose spectra are nearly exactly identical to the original potential.

\vs{1cm}

\pb

\section*{I. Introduction}

Ever since it was conjectured that non-Hermitian Hamiltonians exhibiting
symmetry under the combined transformation of {\it parity} 
($ {\cal{P}}$ : $ x \ra -x $), and {\it time reversal} ($ {\cal{T}} $ : 
$ i \ra -i $) possess a real bound state spectrum \cite{ben}, provided the eigenstates are also simultaneous eigenstates of ${\cal{PT}}$, 
such systems have been widely studied \cite{pt}. 
However, ${\cal{PT}}$
invariance alone is neither necessary nor sufficient to ensure the reality of
the spectrum. It is observed that eigenenergies are 
real for unbroken ${\cal{PT}}$
symmetry, whereas they occur as complex conjugate pairs for the spontaneously
broken case. The latter case corresponds to the case when the Hamiltonian
respects ${\cal{PT}}$ invariance, but eigenfunctions do not. 
Fairly recently it was shown that the 
existence of real or complex
conjugate pairs of energy eigenvalues is attributed to the so-called {\it
pseudo-Hermiticity} of these non-Hermitian Hamiltonians \cite{mostafa} :
\beq
\eta H \eta ^{-1} = H^{\dagger}
\eeq
where $ \eta $ is some Hermitian linear automorphism.
The eigenstates corresponding to real eigenvalues are $ \eta $-orthogonal and
eigenstates corresponding to complex eigenvalues have zero $ \eta $-norm. 
Several non-Hermitian Hamiltonians, whether possessing ${\cal{PT}}$ invariance
or not, have been identified as pseudo-Hermitian under 
$ \eta = e^{- \theta p}$ and $e^{- \phi (x) }$. It is worth mentioning here
that the usual norm, $ \lt< \Psi _n \vert \Psi _n \rt> $, in Hermiticity is 
positive definite, whereas for non-Hermitian Hamiltonians it is indefinite,
being given by
\beq
\lt< \Psi _m \vert \eta \Psi _n \rt> = \eps _n \delta _{m,n}
\eeq
where $\eps = \pm 1$.  

Efforts have always been made to find exactly solvable models 
for the one-dimensional Schr\"{o}dinger equation. Extensive work has been
done on this topic for Hermitian potentials. Similar work has 
been done for complex ${\cal{PT}}$ symmetric potentials as well 
\cite{bb,ikonic,as}. 
The aim of the present work is to generate a series of exactly solvable
non-Hermitian Hamiltonians with real spectra, applying the Darboux
transformation \cite{dar,zheng,rob}. In the Hermitian case, the Darboux transformation is
equivalent to supersymmetric quantum mechanics (SUSYQM). Also,
the method works only when applied to the lowest state. 
For such Hamiltonians, the technique of SUSYQM was generalized to generate
superpotentials by higher excited states \cite{rob}. The domain is split up and the 
partner potentials are defined on specific intervals, depending on the number of singularities. However, for complex
potentials, things are different. Nevertheless, this straightforward method yields an infinite number of exactly
solvable, non-trivial partner potentials, whose spectra are exactly identical
except for the $m ^{th}$ state, provided one starts with the eigenfunction 
$ \psi _m $, corresponding to the $m^{th}$ real eigenvalue $E_m$. Moreover, 
since the new potentials so constructed, do not have any singularity on the real line, they are defined on the entire domain 
$(- \infty, + \infty)$.
As explicit examples, a few non-trivial partners are constructed for two
non-Hermitian potentials with real bound state spectra, viz., \\
(i) ~~~ the ${\cal{PT}}$ symmetric oscillator
\beq
V(x) = (x-i \eps )^2 + \f{\a ^2 - \f{1}{4}}{(x-i \eps )^2}
\eeq
(ii) ~~~ the ${\cal{PT}}$ symmetric version of the generalised Ginocchio potential
\beq
\ba{lcl}  
V(x) &=&
\displaystyle \f{\g ^4 }{ \g ^2 + sinh ^2 u } \lt[ s(s+1) + 1 - \g ^2 -\f{5 \g ^2 (1- \g ^2 )^2}{4( \g ^2 + sinh ^2 u )^2} \rt. \\
&-& \displaystyle \lt. \f{3 (1 - \g ^2 )(3 \g ^2 -1)}{4( \g ^2 + sinh ^2 u )} - 
\lt( \a ^2 - \f{1}{4} \rt) coth ^2 u \rt]
\ea
\eeq
which is an example of an implicit potential, $u$ being a function of $r$ (as
explained in detail in Section IV later on). \\
In each case, the shift from the real axis to the complex plane ensures the removal of singularities from the real line.

\section*{II. Darboux Transformation}

To make this work self contained we first give a brief review of the Darboux
transformation \cite{dar}. 
Though it is applicable to any general 
differential equation, in this section we shall assume the potential to be
real.
We start with a particle moving in the potential 
$V_- (x)$ in the $m^{th}$ state, (i.e., $m$ is the quantum number
equal to the number of nodes, of the $m^{th}$
eigenfunction $ \psi _m (x) $ of the
starting potential $V_- (x)$).
( It must be kept in mind that for complex
potentials with real energies, $m$ denotes the $m^{th}$ energy level; it has
nothing to do with the number of nodes of the eigenfunction, as there are none on the real line. )
If the energy scale is adjusted so that the
$m^{th}$ energy eigenvalue is exactly zero ($ E_m ^{(-)} = 0 $), then the
Schr\"{o}dinger equation reads
\beq
H_- \psi _m  = \lt( - \frac{d^2}{dx^2} + V_- (x) \rt) \psi _m  = 0
\eeq
where the Hamiltonian $H_-$ is given by
\beq
H_- = - \f{d^2}{dx^2} + V_- (x)
\eeq
(The units used are $ \hbar = 2m = 1$ for convenience). \\
Equation (5) has solution
\beq
V_- (x) = \frac{\psi _m ^{\pr \pr}}{ \psi _m}
\eeq 
which is regular everywhere, such that 
\beq
H_-  = \lt( - \frac{d^2}{dx^2} + \frac{\psi _m ^{\pr \pr}}{ \psi _m} \rt) 
\eeq
Thus if the general solution $ \psi = \psi (x) $ 
of the Schr\"{o}dinger equation \cite{rob}
\beq
\f{d^2 \psi }{dx ^2} + \lt[ \eps - V_-(x) \rt] \psi = 0 
\eeq
is known for all values of $ \eps $, and for a particular value of 
$ \eps = E_m $, the particular solution is $ \psi _m $, then the general
solution of the equation 
\beq
\f{d^2 \phi  }{dx ^2} + \lt[ E - V_+(x) \rt] \phi  = 0 
\eeq
with
\beq
V_+(x) = \psi _m (x) \f{d^2}{dx^2} \lt( \f{1}{ \psi _m (x) } \rt)
= 2 \lt( \f{ \psi _m ^{\pr} }{ \psi _m } \rt) ^2 - 
\lt( \f{ \psi _m ^{\pr \pr} }{ \psi _m } \rt)
\eeq
\beq
E = \eps - E_m
\eeq
for $ E \neq 0 $ is
\beq
\ba {lcl}
\phi _n (x) 
&=& \psi _m (x) \lt\{ \f{ \psi _n (x) }{ \psi _m (x) } \rt\} ^{\pr} \\
&=& \psi _n ^{\pr} (x) 
- \lt( \f{ \psi _m ^{\pr} (x) }{ \psi _m (x) } \rt) \psi _n (x) 
\ea
\eeq

Since for Hermitian Hamiltonians the Darboux transformation is equivalent to the intertwining method of SUSY, we shall seek a similar attempt for non-Hermitian Hamiltonians, by 
defining two intertwining operators $A$ and $B$ :
\beq
A = \frac{d}{dx} + W_m
\eeq
\beq
B = - \frac{d}{dx} + W_m
\eeq
where
\beq
W_m = - \f{ \psi _m ^{\pr} }{ \psi _m}
\eeq
Then 
\beq
H_- = BA = \lt( - \frac{d^2}{dx^2} + V_- (x) \rt) 
\eeq
Now let us construct a partner Hamiltonian $H_+$ by
\beq
H_+ = AB = \lt( - \frac{d^2}{dx^2} + V_+ (x) \rt) 
\eeq
where
\beq
V_+ = V_- - 2 W_m
\eeq
such that
\beq
V_{\pm} = W _m ^2 \pm W _m ^{\pr}
\eeq
Evidently if $ \psi _n $ is an eigenfunction of $H_-$ with
energy eigenvalue $E_n ^-$, then $ \phi _n = A \psi _n $ is also an 
eigenfunction of $H_+$ with the same energy eigenvalue $E_n ^-$, 
for specific values of $n$.
\beq
H_+ A \psi _n = (AB)A \psi _n = A(H_- \psi _n ) = E_n ^- (A \psi _n )
\eeq
Thus, in case of Hermitian SUSY QM, the potentials 
$V_+$ and $V_-$ are isospectral except for the $m$ lowest states of 
$V_-$, for which there is no corresponding state of $V_+$. Consequently, the
ground state of $V_+$ is $E_0 ^+ = E_{m} ^- $. All higher states have
identical energies. Furthermore, $W_m $ is the superpotential, and 
$ B = A^{\dagger} $. $ V_{\pm}$ are called SUSY-$m$ partner potentials.
Since $ \psi _m ^{\pr} \neq 0 $ at $x_j$, $W_m $ has singularities at the nodes $x_j$, $j=1,2,3, \cdots $ of $ \psi _m $. 
For $m=0$, the usual SUSY partners are defined on $(- \infty, + \infty )$, 
for $m=1$, there are two separated potential wells, each of them on a
semi-infinite domain, for $m=2$ there is one infinite potential well on a finite domain between nodes $x_1$ and $x_2$, and two binding potential 
wells on the two semi-infinite domains $(- \infty, x_1]$ and 
$[x_2, + \infty )$, and so on \cite{rob}. 

However, the scenario is different in case of non-Hermitian quantum mecahnics. If there exists a linear, invertible Hermitian operator $ \eta $ such that \cite{mostafa}
\beq
B = A^{\#} = \eta ^{-1} A ^{\dagger} \eta
\eeq
then the partner Hamiltonians can be written as 
\beq
H_+ = A A^{\#} \ \ \ \ \ \ \ \ H_- = A^{\#} A 
\eeq 
and $A$ and $B$ are mutual pseudo-adjoints, so that $V_{\pm}$ can be termed as pseudo-supersymmetric partners. Moreover, the new potentials are defined on the entire domain
as the complex potentials have no singularities on the real axis.
Thus, if one of the partner systems is exactly solvable, this SUSY induced
formalism enables one to solve the other non-trivial partner as well. \\
It may be worth mentioning here that if $V_{\pm}$ are
isospectral, so are 
\beq
v_{\pm} = V_{\pm} - \beta _m
\eeq
where, $ \beta _m $ is some $n$-independent arbitrary constant (real or imaginary). The last expression ensures the eigenspectrum to be real under ${\cal{PT}}$ invariance as explained below. If the eigen energies of the complex potential 
$V_{\pm} (x) $ are 
$ E_n ^{\pm} + \b _m $, where $ E_n ^{\pm}$ are 
some $n$-dependent real constants, and $ \b _m $ are as defined above, then 
the eigen energies of $v_{\pm} (x) $ are 
$E_n ^{\pm}$. Thus eq. $(24)$ ensures the eigenspectrum 
to be real under ${\cal{PT}}$ invariance, provided the complex potentials
$v_{\pm}$ are ${\cal{PT}}$ symmetric .

\section*{III. ${\cal{PT}}$ symmetric oscillator}

We start with the widely studied ${\cal{PT}}$ symmetric oscillator 
\cite{znojil}
\beq
V(x) = z^2 + \f{\a ^2 - \f{1}{4}}{z^2}
\eeq
\beq
z = x - i \eps
\eeq
which is known to possess the double set of eigenfunctions 
\beq
\psi _{nq} = N_{nq} e^{ - \f{ z^2}{2} } z^{-q \a + \f{1}{2} } L_n ^{
(- q \a )} \lt( z^2 \rt) 
\eeq
with eigenvalues
\beq
E_{nq} = 4n + 2 - 2 q \a
\eeq
where $ L_n ^{( \sigma )} $ are the associated Laguerre polynomials  given by \cite{hand}
\beq
L_n ^{(\sigma)} (y) = \f{\Gamma \lt( n+ \sigma + 1 \rt)}{
\Gamma \lt( n+1 \rt) ~ \Gamma \lt( \a + 1 \rt) } M \lt( -n, \sigma +1, y \rt)
\eeq
and $q = \pm 1$ is the {\it quasi-parity}.\\
Starting with the eigenfunction $ \psi _{mq} $, corresponding to the 
$m^{th}$ eigen value which is real, the 
pseudo-superpotential term $ W _{mq} $  is
calculated to be
\beq
\ba {lcl}
\displaystyle W _{mq} 
&=& \displaystyle - \f{ \psi _{mq} ^{\pr}}{ \psi _{mq}} \\
&=& \displaystyle -z + \f{ \lt( -q \a + 2m + 3/2 \rt) }{z}  - \f{ 2 \lt( m+1 \rt)}{z}
\f{L_{m+1} ^{(-q \a)} \lt( z^2 \rt)}{L_m ^{(-q \a)} \lt( z^2 \rt) }
\ea
\eeq
for $m=0,1,2, \cdots $, where prime 
denotes differentiation with respect to $x$. \\
The isospectral partner potentials for various values of $m$ 
are obtained from the formula
\beq
v_{+q} ^{(m)} (x) = W _{mq} ^2 + W _{mq} ^{\pr} - \beta _{mq}
\eeq
with 
\beq
\beta _{mq} = 2q \a - 2 \lt( 2m +1 \rt)
\eeq
The potential $v_{+q} ^{(m)} (x) $ shares all the energy levels of 
the (energy shifted) ${\cal{PT}}$ symmetric oscillator
given in eq.(3), with $ V(x) = v_- (x) = W _{mq} ^2 - W _{mq} ^{\pr} 
- \beta _{mq} $, except for the $m ^{th}$ state of $v_-$ which has no
counterpart in $v_+$. 
Using the explicit forms of the Laguerre polynomials \cite{hand},
the partners for $m=0,1,2$ are obtained below :

\vs{.5cm}

\noindent
~~(i) ~~~ $m=0$
\beq
\displaystyle v_{+q} ^{(0)} = z^2 + \f{ \a ^2 - 2q \a + 3/4}{z^2} + 2
\eeq
This is analogous to the Hermitian case of a satellite potential.

\vs{.5cm}

\noindent
~(ii) ~~~ $ m=1 $ \\
This gives the first non-trivial partner.
\beq
v_{+q} ^{(1)} = z^2 + \f{ \a ^2 - 2q \a + 3/4}{z^2} + \f{4}{(-q \a +1)-z^2}
+ \f{8z^2}{ \lt\{ (-q \a +1) -z^2 \rt\} ^2 } + 2
\eeq
with ground state (GS) wave function 
\beq
\phi _{0q} = \f{2}{-q \a +1 - z ^2} e^{- ~ \f{z^2}{2} } z^{- q \a +
\f{3}{2}} 
\eeq
and GS energy
\beq
E_{0q} = 2 - 2q \a
\eeq
It can be shown that because of normalization criterion, 
$ \phi _{1q} $ has to be excluded from the spectrum. 
All other states $(n=1,2,3 \cdots)$ can be obtained from (13) as
\beq
\phi _{(n+1)q} (x) = \f{f_1 ~ f_{(n+1)} ^{\pr} - f_1 ^{\pr} ~ 
f_{(n+1)}}{f_1} \psi _0
\eeq
where $f_n$ stands for $L^{-q \a} _n \lt( z^2 \rt) $, and prime 
denotes differentiation with respect to $x$. 
Since the functions $L^{-q \a} _n \lt( z^2 \rt) $ are well defined over the
entire real line, singularities do not appear for $ \phi  _{nq} (x) $. 
Also, the eigenfunctions are well behaved and normalizable.

\vs{.5cm}

\noindent
(iii) ~~ Simlarly for $ m = 2 $ 
\beq
v_{+q} ^{(2)} = z^2 + \f{ \a ^2 - 2q \a + 3/4}{z^2} - 
\f{4 \lt[ 3z^2 - (-q \a +2) \rt] }{L_2 ^{-q \a} \lt( z^2 \rt)}
+ \f{8z^2 \lt[ z^2 -(-q \a +2) \rt] ^2}{ \lt\{L_2 ^{-q \a} \lt( z^2 \rt) \rt\}} + 2
\eeq
As is observed, there are two partner potentials, characterized by the
quasi-parity $q = \pm 1$, for each $m$ value. Also, the ${\cal{PT}}$ 
invariance of the partner potentials depends on the
parameter $\a$. For real $ \a $, the partners retain their ${\cal{PT}}$
symmetry, whereas for imaginary $ \a $, we obtain new 
complex potentials with real energies. Moreover, since 
there are no singularities on the real axis, 
the new potentials so constructed are defined on 
the entire domain $(- \infty, + \infty)$.
We have plotted the real and imaginary parts of the partner potential 
$ v_{+q} ^{(1)} $ (34) and the ground state wave function $ \phi _{0q}$
(35) of the same in Fig. 1 and 2 respectively.

\section*{IV. The generalised Ginocchio potential }

Our next attempt is to find the isospectral partners of the generalized
Ginocchio potential given by \cite{gin}
\beq
\ba{lcl}
\displaystyle V(r) 
&=& \displaystyle \f{\g ^4 }{ \g ^2 + sinh ^2 u } \lt[ s(s+1) + 1 - \g ^2 
-\f{5 \g ^2 (1- \g ^2 )^2}{4( \g ^2 + sinh ^2 u )^2} \rt. \\
&-& \displaystyle \lt. \f{3 (1 - \g ^2 )(3 \g ^2 -1)}{4( \g ^2 + sinh ^2 u )} - 
\lt( \a ^2 - \f{1}{4} \rt) coth ^2 u \rt]
\ea
\eeq
This is an example of an implicit potential, as it is expressed in terms of a
function $u(r)$ which is known only in the implicit form :
\beq
\ba{lcl}
\displaystyle r 
&=& \displaystyle \f{1}{ \g ^2} \lt[ tanh ^{-1} \lt\{ \lt( \g ^2 + sinh ^2 u \rt) 
^{-\f{1}{2}} sinh ~u \rt\} \rt. \\
&+& \displaystyle \lt. \g ^2 -1 ) ^{\f{1}{2}} tan ^{-1} \lt\{ ( \g ^2 - 1 ) ^{\f{1}{2}}
( \g ^2 + sinh ^2 u )^{- \f{1}{2}} sinh ~u \rt\} \rt]
\ea
\eeq
The monotonously increasing function $u(r)$ is the solution of the first order
differential equation
\beq
\f{du}{dr} = \f{ \g ^2 cosh ~u }{( \g ^2 + sinh ^2 u )^{\f{1}{2}}}
\eeq
The bound state eigenfunctions are expressed in terms of the Jacobi
polynomials as
\beq
\psi _0 ^{(n)} (r) = N_n ( \g ^2 + sinh ^2 u )^{\f{1}{4}} 
\lt( sinh ~u \rt) ^{-q \a + 1/2} \lt(cosh ~u \rt) ^{- \m _n +q \a -1} f_n
\eeq
where 
\beq
f_n = P_n ^{( \m _n, -q \a  )} \lt( 2 ~ tanh ^2 u - 1 \rt)
\eeq
are the Jacobi polynomials given by \cite{hand}
\beq
{\cal{P}}_n ^{\a,\b} (z) =  \f{\Gamma \lt( n+ \sigma + 1 \rt)}{
\Gamma \lt( n+1 \rt) ~ \Gamma \lt( \a + 1 \rt) } ~ 
F \lt( -n, n + \a + \b + 1, \a + 1, \f{1-z}{2} \rt)
\eeq
and $N_n$ are normalization constants. The bound states are located at
\beq
E_n = - \g ^4 \m _n ^2 , ~~~~~~~~~~~~~~~~~~~~~ n = 0, 1, 2, \cdots < \f{1}{2}
\lt( s + q \a - \f{1}{2} \rt)
\eeq
with
\beq
\m _n = \f{1}{ \g ^2} \lt[ - \lt( 2n -q \a + 1 \rt)
+ \lt\{ \lt( 2n - q \a + 1 \rt) ^2 
\lt( 1 - \g ^2 \rt) + \g ^2 \lt( s + \f{1}{2} \rt) ^2 \rt\} ^{\f{1}{2}} \rt]
\eeq
Analogous to the ${\cal{PT}}$ symmetric oscillator, the presence of
quasi-parity ($q = \pm 1$) gives rise to a double set of solutions. \\
Starting with the eigenfunction, $ \psi _{mq} $, 
corresponding to real eigenvalue $E_m$,  we obtain
\beq
\ba {lcl}
\displaystyle W_{mq} 
&=& \displaystyle - \f{ \psi _{mq} ^{\pr}}{ \psi _{mq}} \\
&=& \displaystyle - \f{ \gamma ~ ^2  sinh ~ u ~~ cosh ^2 u}{ 2 ( \gamma ~ ^2 + sinh ^2 u) ^{\f{3}{2}} } \\
&+& \displaystyle \f{ \gamma ^2 }{ ( \gamma ^2 + sinh ^2 u ) ^ {\f{1}{2}} } 
\lt \{ \lt( \mu _m + \f{1}{2} \rt) sinh ~ u + \f{ \lt( q \alpha - 
\f{1}{2} \rt ) }{ sinh ~ u} \rt \} - \f{f_m ^{\pr}}{f_m}
\ea
\eeq
for $m=0,1,2, \cdots $, where prime 
denotes differentiation with respect to $x$. \\
Proceeding in a manner similar to that shown above, one can obtain new,
exactly solvable, ${\cal{PT}}$ symmetric potentials which share the same eigen energies as the generalised
Ginocchio potential given in (39), with the help of 
\beq
v_{+q} ^{(m)} (x) =  W_{mq}  ^2 +  W_{mq}  ^{\pr} - \beta _{mq}
\eeq
where 
\beq
\beta _{mq} = \g ^4 \m _m ^2
\eeq
As an explicit example, we give the results for $m=1$ only. 


$$ v_{+q} ^{(1)} (x) = - \f{ \g ^4}{ \g ^2 + sinh ^2 u } \lt \{ s(s+1) 
+ \g ^2 ( 1 - 2 \m _1 ) + 2 q \a - 2 \rt. $$
$$  - \lt ( \a ^2 - 2q \a + \f{3}{4} \rt ) coth ^2 u
+ \f{7}{4} ~  \f{ \g ^2 (1 - \g ^2 ) ^2 }{ (\g ^2 + sinh ^2 u ) ^2} $$
$$ \lt. -  \f{ ( 1 - \g ^2)}{ \g ^2 + sinh ^2 u } \lt [ \g ^2 \lt( 2 \m _1
- \f{11}{4} \rt) + \f{9}{4} - 2q \a \rt ] \rt\} $$
$$ -4 \g ^4 \f{ \lt( \m _1 - q \a + 2 \rt) }{ \g ^2 + sinh ^2 u }
+ \f{4 \g ^4 }{ (f_1)^2} \f{ \lt( \m _1 - q \a + 2 \rt)^2 
}{\lt( \g ^2 + sinh ^2 u \rt) } tanh ^2 u sech ^2 u $$
\beq
- \f{4 \g ^4 }{ f_1} \f{ \lt( \m _1 - q \a + 2 \rt) }{ \g ^2 + sinh ^2 u }
\lt[ -2 tanh ^2 u + \f{ \g ^2}{  \g ^2 + sinh ^2 u } \rt] 
\eeq
where
\beq
f_1 = q \a - 1 + \lt( \m _1 - q \a + 2 \rt) ~ tanh ^2 u 
\eeq
Thus one obtains two 
isospectral partners for each $m$, because of the presence of quasi-parity. 
If $\a $ is real, then the new potentials thus formed are ${\cal{PT}}$
symmetric. However, for $ \a $ pure imaginary, we get non-Hermitian
non-${\cal{PT}}$ invariant potentials with real bound state spectra.
Since neither $f_1$ or $ \lt( \g ^2 + sinh ^2 u \rt) $ has any root on
the real axis ($u$ being complex), the new potential so constructed is defined
on the entire domain $(- \infty, + \infty)$, in this case as well.

\section*{V. Conclusions }

To conclude, starting with the eigenfunction $ \psi _m$ corresponding to the 
$m^{th}$ real eigenvalue $E_m$, of the potential $v_- (x)$, new, exactly solvable, non-trivial
partners have been constructed for two ${\cal{PT}}$ symmetric potentials,
viz., \\
(i) ~~~ the ${\cal{PT}}$ symmetric oscillator, \\
(ii) ~~ the ${\cal{PT}}$ symmetric version of the generalised Ginocchio potential \\
the partners being related to the original potential $v_-(x)$ by
$$ v_{\pm}  =  W_m ^2  \pm W_m  - \beta _m $$
where $\b _m$ is some $n$-independent constant and 
$$ W_m = - \f{ \psi _m ^{\pr}}{ \psi _m} $$
Though $m=0$ case gives the usual shape-invariant form, higher  values 
of $m$ generate new examples of non-shape invariant partners, 
as they do not obey the shape-invariance condition \cite{gend},
which states that if the profiles of $V_{\pm}$ are such that they satisfy the  relationship
\beq
V_- (x,a_0) = V_+ (x,a_1) + R(a_1)
\eeq
where $a_1$ is some functon of $a_0$ (say, $a_1 = f(a_0)$), then only the potentials $V_{\pm}$ are termed as shape-invariant. 
However, the new exactly solvable potentials constructed in this work do not fall in this category.
Moreover, the new potentials admit all the 
energies of the original potential except for the
$m^{th}$ state, which is missing in the partner potentials, unlike the absence of the lowest $m$ states in the Hermitian case. Furthermore,
contrary to the Hermitian case where the partners are defined on specific disjoint intervals depending on the level $m$, in this case the partners are defined on the entire domain 
$( - \infty, + \infty)$, as there are no singularities on the real line.

It will be interesting to investigate the effect of 
spontaneous ${\cal{PT}}$ symmetry breaking on such partners. We propose to
take up this study in the recent future.

\section*{Acknowledgment}
One of the authors (A.S.) is grateful to the Council of Scientific \& Industrial Research,
India, for granting her financial assistance.

\newpage

\noindent
{\bf Figure Captions}

\vs{1cm}

\noindent
{\bf Fig. 1 :} Graph of the real (dotted curve) and imaginary (solid curve) 
parts of the potential $ v_{+q} ^{(1)}$ given in eq.(34) for $ \a = 3/4, \ q = +1, \ \eps = 1 $.

\vs{1cm}

\noindent
{\bf Fig. 2 :} Graph of the real (dotted curve) and imaginary (solid curve) 
parts of the wave function $ \phi _{0q}$ given in eq.(35) for $ \a = 3/4, \ q = +1, \ \eps = 1 $.

\ed